# A Novel Steganography Algorithm for Hiding Text in Image using Five Modulus Method

Firas A. Jassim
Management Information Systems Department,
Faculty of Administrative Sciences,
Irbid National University,
Irbid  2600, Jordan

## ABSTRACT
The needs for steganographic techniques for hiding secret message inside images have been arise. This paper is to create a practical steganographic implementation to hide text inside grey scale images. The secret message is hidden inside the cover image using Five Modulus Method. The novel algorithm is called (ST-FMM. FMM which consists of transforming all the pixels within the 5×5 window size into its corresponding multiples of 5. After that, the secret message is hidden inside the 5×5 window as a non-multiples of 5. Since the modulus of non-multiples of 5 are 1,2,3 and 4, therefore; if the reminder is one of these, then this pixel represents a secret character. The secret key that has to be sent is the window size. The main advantage of this novel algorithm is to keep the size of the cover image constant while the secret message increased in size. Peak signal-to-noise ratio is captured for each of the images tested. Based on the PSNR value of each images, the stego image has high PSNR value. Hence this new steganography algorithm is very efficient to hide the data inside the image.

## 1. INTRODUCTION
Security of information becomes one of the most important factors of information technology and communication because of the huge rise of the World Wide Web and the copyrights laws. Cryptography was originated as a technique for securing the confidentiality of information. Unfortunately, it is sometimes not enough to keep the contents of a message secret, it may also be necessary to keep the existence of the message secret and the concept responsible for this is called steganography [5]. Steganography is the practice of hiding secret message within any media. Most data hiding systems take advantage of human perceptual weaknesses. Steganography is often confused with cryptography because the two are similar in the way that they both are used to protect secret information. If both the techniques: cryptography and steganography is used then the communication becomes double secured [19].The main difference between Steganography and cryptography is that, cryptography concentrates on keeping the contents of a message secret while steganography concentrates on keeping the existence of a message secret [20]. Steganography and cryptography are both needed to protect messages from third party but each one with its own. Thus, when there is a need protect the presence of message; the steganography is the solution [20]. Probably most common cover media are multimedia objects which are images, audio, and video. Here, in this paper, we focus on images as cover media. Two other technologies that are closely related to steganography are watermarking and fingerprinting [2]. These technologies are mainly concerned with the protection of intellectual property. Examples of common application of steganography are in the field of copyright protection. According to [17], the information hidden in the bit stream allows an early resynchronization of the video. The only price to pay is a small degradation of the undamaged video quality, with a very limited increase in computational complexity [15].

Steganographic technique finds its main application in the field of secret communication. It can be used by intelligence agencies across the world to barter highly confidential data in a secret media, e.g. a secret agent can hide a map of a terrorist camp in a photograph using image steganographic software and post it on a forum. An officer from the head office could download the photograph from the forum and easily retrieve the hidden map [11].

The outline of the paper is as follows: An overview of image steganography was reviewed in Section 2. Five Modulus Method (FMM) was discussed in Section 3. The proposed steganography algorithm was presented in Section 4. Experimental results and conclusions are presented in Sections 5 and 6, respectively.

## 2. STEGANOGRAPHY PRELIMINARIES
In the last decade, it has been an increasing interest in using images as cover media for steganographic communication. The word steganography is originally derived from Greek words, which mean "Covered Writing". It has been used in various forms for thousands of years. In the fifth century BC Histaiacus shaved the head of his messenger, wrote the message on his scalp, and then waited for the hair to grow again. The messenger, clearly carrying nothing pugnacious, could travel freely. Arriving at his destination, he shaved his head and the secret message could be easily read by the receiver [13][14]. One of the oldest methods to hide a message inside a text is to take the first letter of each word. To illustrate this, suppose the following sentence 'Since everyone can read, encoding text in neutral sentences is doubtfully effective'. By taking the first letter of each word we get the secret message which is 'Secret inside' [19]. Even later, the Germans developed a technique called the microdot. Microdots are photographs with the size of a printed period but contain full page information. The microdots where then printed in a letter or on an envelope and being so small, they could be sent unnoticeable [10].

However, steganography has its place in security. Though it cannot replace cryptography totally, it is intended to append it. Steganography can be used along with cryptography to make a highly secure data. It is not the same as watermarking [8]. The key difference between steganography and watermarking is the absence of an opponent. In watermarking applications like copyright protection and authentication, there is an active opponent that would attempt to remove, abolish or forge watermarks. In steganography there is no such active opponent as there is no value associated with the act of removing the information hidden in the content [11]. According to [8], two types of Steganography were categorised. The first on which is called fragile, this steganography involves embedding information into a file which is destroyed if the file is modified. On the other hand, the other type is the robust steganography which aims to embed information into a file that cannot be easily destroyed. The best known steganographic method that works in the spatial domain is the Least Significant Bit (LSB) which replaces the least significant bits of pixels selected to hide the information. LSB is a one of the widest and simplest methods used in image steganography. Data hidden in images using this method is highly sensitive to image alteration and vulnerable to attack. A detailed discussion about LSB could be found in [1][4][6][23]. Also, there are a wide variety of different techniques with their own advantages and disadvantages were constructed in steganography. Research in hiding data inside image using steganography technique has been done by many researchers [7][9][16][21][22]. Also, an excellent theoretical background about steganography could be found in [3].

## 3. FIVE MODULUS METHOD

The Five Modulus Method (FMM) was firstly proposed by [12]. The fundamental idea behind FMM is based upon the following concept: A common characteristic in most of images is that the neighbouring pixels are correlated. Therefore, for bi-level images, the neighbours of a pixel tend to be similar to the original pixel. Hence, FMM consists of dividing the image into blocks of k×k pixels each. Clearly, in bi-level grey images, we know that each pixel is a number between 0 and 255. Therefore, if we can transform each number in that range into a number divisible by 5, then this will not affect the Human Visual System (HVS). The basic idea in FMM is to check the whole pixels in the k×k block and transform each pixel into a number divisible by 5 according to the following algorithm.

> If Pixel mod 5 = 4
> Pixel=Pixel+1
> Else if Pixel mod 5 = 3
> Pixel=Pixel+2
> Else if Pixel mod 5 = 2
> Pixel=Pixel-2
> Else if Pixel mod 5 = 1
> Pixel=Pixel-1

where Pixel is the digital image representation of the k×k block. According to table (1), the transformation of the FMM could be demonstrated.

**Table 1 FMM transformation**

| Old | New | | Old | New |
|-----|-----|---|-----|-----|
| 0 | 0 | | 111 | 110 |
| 1 | 0 | | 112 | 110 |
| 2 | 0 | | 113 | 115 |
| 3 | 5 | | 114 | 115 |
| 4 | 5 | | 115 | 115 |
| 5 | 5 | … | … | … |
| 6 | 5 | | 221 | 220 |
| 7 | 5 | | 222 | 220 |
| 8 | 10 | | … | … |
| 9 | 10 | | 254 | 255 |
| 10 | 10 | | 255 | 255 |

Here, FMM could transform any number in the range 0-255 into a number that when divided by 5 the reminder is always 0, 1, 2, 3, or 4, (e.g., 20 mod 5 is 0, 11 mod 5 is 1, 202 mod 5 is 2, 188 mod 5 is 3 and so on). Mathematically speaking, the new values for the k×k block will always be as follows: 0,5,10,15,20,25,30,35,40,…,200, 205,210,215,...,250, 255, (i.e. multiples of 5).

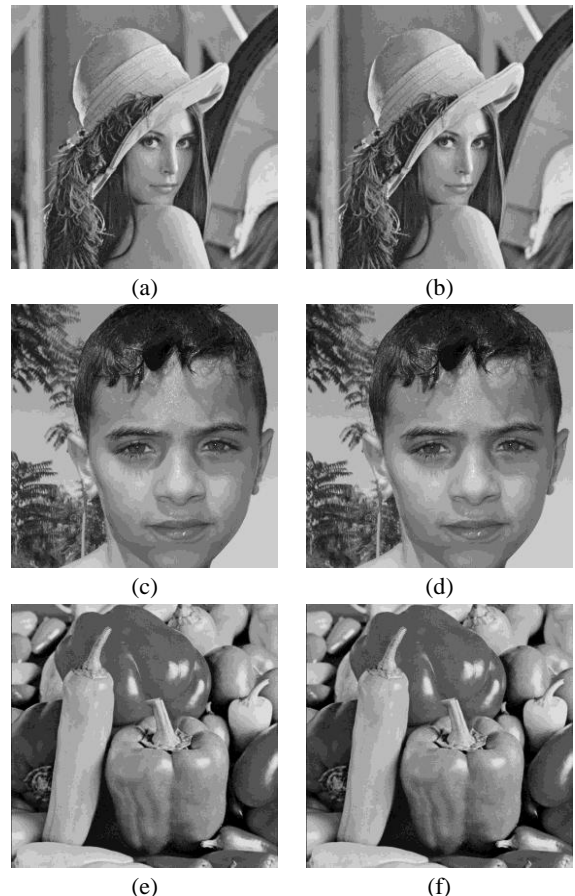

(a)     (b)

(c)     (d)

(e)     (f)

**Figure 1. Original images (left) and their FMM transformation (right)**

From figure (1), we can see that the human eye can not differentiate between the original images and the transformed FMM images. In addition, to support our claim the PSNR values were calculated for the test images after the FMM transform. The result could be shown clearly in table (2).

**Table 2. PSNR values for FMM images**

| FMM |
|-----|

| Lena | 51.0611 |
|---|---|
| Saif | 35.2874 |
| Peppers | 47.2951 |

## 4. PROPOSED STEGANOGRAPHY ALGORITHM

According to the previous section, the FMM transformation does not affect the Human Visual System (HVS). The proposed algorithm was called ST-FMM which means STeganography by the Five Modulus Method. Therefore, all the pixels inside the FMM images are all multiples of 5 only. Hence, the values that are not divisible by 5 are distinct inside k×k block. Obviously, it is known that the standard ASCII code consists of 128 characters. But the most 95 common characters used in binary coding could be extracted from the general ASCII code and represented in table (3).

**Table 3. The most 95 common ASCII characters**

| Dec | Char | Dec | Char | Dec | Char | Dec | Char | Dec | Char |
|---|---|---|---|---|---|---|---|---|---|
| 32 | space | 52 | 4 | 72 | H | 92 | \ | 112 | p |
| 33 | ! | 53 | 5 | 73 | I | 93 | ] | 113 | q |
| 34 | " | 54 | 6 | 74 | J | 94 | ^ | 114 | r |
| 35 | # | 55 | 7 | 75 | K | 95 | _ | 115 | s |
| 36 | $ | 56 | 8 | 76 | L | 96 | ` | 116 | t |
| 37 | % | 57 | 9 | 77 | M | 97 | a | 117 | u |
| 38 | & | 58 | : | 78 | N | 98 | b | 118 | v |
| 39 | ' | 59 | ; | 79 | O | 99 | c | 119 | w |
| 40 | ( | 60 | < | 80 | P | 100 | d | 120 | x |
| 41 | ) | 61 | = | 81 | Q | 101 | e | 121 | y |
| 42 | * | 62 | > | 82 | R | 102 | f | 122 | z |
| 43 | + | 63 | ? | 83 | S | 103 | g | 123 | { |
| 44 | , | 64 | @ | 84 | T | 104 | h | 124 | \| |
| 45 | - | 65 | A | 85 | U | 105 | i | 125 | } |
| 46 | . | 66 | B | 86 | V | 106 | j | 126 | ~ |
| 47 | / | 67 | C | 87 | W | 107 | k | | |
| 48 | 0 | 68 | D | 88 | X | 108 | l | | |
| 49 | 1 | 69 | E | 89 | Y | 109 | m | | |
| 50 | 2 | 70 | F | 90 | Z | 110 | n | | |
| 51 | 3 | 71 | G | 91 | [ | 111 | o | | |

### 4.1 Determination of Window Size

The determination of the suitable window size used for steganography is very important procedure. The smaller window size is better to increase number of secret message characters hidden in the cover image. In this article, a general formula to determine the suitable window size has been derived as follows:

$$Window\ size = \left\lceil \sqrt{\left\lceil \frac{n}{4} \right\rceil} \right\rceil \quad (1)$$

where n represents the number of distinct characters used in the secret message text. The $\lceil . \rceil$ operator used as a ceiling function to approximate the floating number into the nearest upper integer. The number of values inside the k×k window is k2. Therefore, to increase the number of characters to be accommodated inside the k×k window, a loop procedure was innovated. As mentioned previously, the reminders of 5 which are non-multiples of 5 are 1, 2, 3, and 4. If the reminder is 1, this means that we are in first loop, i.e. within the same k×k window. If the reminder is 2, this means that we are in the second loop of the k×k window, and so on. Hence, the exact number of values that may be accommodated within the k×k window size is:

$$No.\ of\ values\ within\ one\ window = 4k^2 \quad (2)$$

Also, a general formula to extract the character ASCII value from the steganography image has been derived as follows:

$$Character\ value = (position + (re\min der - 1) \times K^2) + (starting\ index - 1) \quad (3)$$

### 4.2 Case of 5×5 window Size

By considering table (3), and substituting the value of n by 3 in eq.(1), this yield.

$$Window\ size = \left\lceil \sqrt{\left\lceil \frac{95}{4} \right\rceil} \right\rceil = 5$$

Hence, the suitable window size that could be used to represent the most frequent ASCII characters is 5. Since the goal in this paper is to reduce the window size to accommodate more hidden text. Therefore, a 5×5 window size will be used. According to eq.(2), this will accommodate 4×52=100 value within the 5×5 window size. Generally, to use all the 128 characters in ASCII table, one can substitute n by 128 in eq.(1) to get window size equals 6.

Now, an illustrative example will discussed using 5×5 window size to hide a secret message 'A Steganography Algorithm for Hiding Text in Image Using Five Modulus Method.' in the cover image (peppers.bmp), this could be shown in table (4).

**Table 4. 5×5 stego-peppers bitmap image**

| 35 | 70 | 60 | 65 | 65 | 61 | 50 | 55 | 55 | 60 |
|---|---|---|---|---|---|---|---|---|---|
| 50 | 115 | 110 | 110 | 110 | 120 | 105 | 105 | 110 | 110 |
| 35 | 115 | 120 | 110 | 110 | 125 | 115 | 105 | 110 | 110 |
| 35 | 117 | 115 | 110 | 105 | 115 | 105 | 100 | 105 | 105 |
| 50 | 120 | 120 | 120 | 115 | 115 | 110 | 120 | 115 | 115 |

| 65 | 70 | 75 | 80 | 80 | 80 | 90 | 85 | 85 | 85 |
|---|---|---|---|---|---|---|---|---|---|
| 113 | 110 | 110 | 110 | 105 | 105 | 105 | 100 | 100 | 105 |
| 110 | 110 | 110 | 110 | 105 | 105 | 110 | 110 | 110 | 110 |
| 105 | 110 | 110 | 110 | 110 | 105 | 115 | 110 | 105 | 105 |
| 115 | 115 | 115 | 110 | 110 | 105 | 124 | 115 | 110 | 110 |

According to table (4), it is clearly shown that a 5×5 window size was implemented. The first 5×5 window size contains all elements inside are modulus of 5 except 117 and 117 mod 5 gives 2 which means two loops. Also, the position of 117 in the first window is 9 by column-wise tracing. According to eq.(3), the value of the hidden character is computed as: (9 +1*25) + 31 = 65 which is ASCII code for A. The value of 25 was used as the square of the window size, 52=25. Moreover, since the starting decimal character of the 95 most frequent ASCII characters is 32, to make the numbering starting from 1, the number of 31 was used. A similar procedure may be applied to the residue windows as follows. Therefore, for $61 \Rightarrow (1+0*25) + 31 = 32$ which is ASCII code for space. Also, $113 \Rightarrow (2 + 31 + 2*25) = 83$ which is ASCII code for S. Finally, $124 \Rightarrow (10 + 31 + 3*25) = 116$ which is ASCII code for t.

### 4.3 Case of 3×3 Window Size

Someone need to hide only a text in a cover image without numbers and special characters and this text either be in an upper case or lower case. This is does not matter because the receiver on the other side needs just an information which may be upper or lower. Therefore, the 26 alphabet characters could be used to construct a special window size just for English letters. According to eq.(1), we can substitute n by 26 to get:

$$Window\,size = \left\lceil \sqrt{\left\lceil \frac{26}{4} \right\rceil} \right\rceil = 3$$

Obviously, 3×3 window size is smaller than 5×5 window size and this implies more text to be hidden in the cover image. Now, an illustrative example will be discussed using 3×3 window size to hide a secret message 'to be or not to be' in the cover image (Saif.bmp), this could be shown in table (5).

**Table 5. 3×3 stego-Saif bitmap image**

| 50 | 45 | 45 | 45 | 45 | 45 | 55 | 55 | 60 | 65 | 75 | 90 |
|----|----|----|----|----|----|----|----|----|----|----|----|
| 53 | 45 | 40 | 40 | 35 | 35 | 40 | 44 | 45 | 56 | 85 | 100 |
| 45 | 40 | 35 | 35 | 37 | 30 | 30 | 35 | 35 | 55 | 80 | 90 |

| 95 | 100 | 115 | 120 | 95 | 80 | 65 | 60 | 55 | 65 | 120 | 160 |
|----|-----|-----|-----|----|----|----|----|----|----|-----|-----|
| 110 | 121 | 130 | 95 | 84 | 70 | 55 | 50 | 50 | 45 | 95 | 140 |
| 95 | 100 | 85 | 75 | 65 | 50 | 45 | 42 | 40 | 40 | 65 | 132 |

Now, to retrieve the hidden text in table (5), the first 3×3 window contains 53 as non-multiples of 5. Hence, according to eq.(3), we get:

$53 \Rightarrow (2+2*9) + 96 = 116$ which is ASCII code for t.

$37 \Rightarrow (6+1*9) + 96 = 111$ which is ASCII code for o.
$44 \Rightarrow 44 \bmod 5 = 4$ which is ASCII code for space.
$56 \Rightarrow (2+0*9) + 96 = 98$ which is ASCII code for b.
$121 \Rightarrow (5+0*9) + 96 = 101$ which is ASCII code for e.
$84 \Rightarrow 84 \bmod 5 = 4$ which is ASCII code for space.
$42 \Rightarrow (6+1*9) + 96 = 111$ which is ASCII code for o.
$132 \Rightarrow (9+1*9) + 96 = 112$ which is ASCII code for r.

where the exact hidden text is 'to be or not to be'.

The value of 9 was used as the square of the window size, 3²=9. Moreover, since the starting decimal character of the small ASCII characters is 97, to make the numbering starting from 1, the number of 96 was used.

### 4.4 Secret stego-key

A stego-key is used to control the hiding process so as to restrict detection and recovery of the secret message [18]. In ST-FMM, the secret key is used for extracting the secret message from the cover image is the window size. Therefore, discovering the secret message from the receiver relies solely on knowing of the window size. If an intruder intended to extract the secret message from the cover image then a tremendous number of possibilities must be attempted. In private key steganography both the sender and the receiver share a secret key which is used to embed the message [5]. It must be mentioned that, if the most 95 common ASCII characters were used then there is no need to use any secret key because it is known the window size is 5.

## 5. EXPERIMENTAL RESULTS

In order to demonstrate the proposed steganography algorithm, ST-FMM has been implemented to three bitmap test images (Lena, Saif, and Peppers) which are used as a cover images. All of the test images are 512×512 bitmap images. Also, six text files have been used as a secret message which has to be hidden inside the cover images. Since the text files used contain the most 95 common ASCII characters, a window of size 5×5 have been used to adopt ST-FMM. The quality of the stego images have been measured using PSNR (Peak signal-to-noise ratio). PSNR is a standard measurement used in steganography technique in order to test the quality of the stego images. The higher the value of PSNR, the more quality the stego image will have.

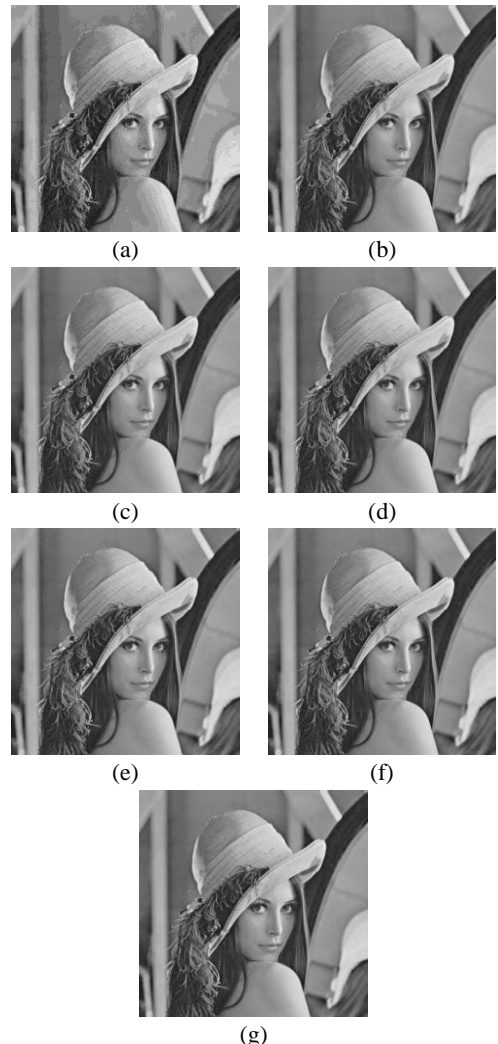

**Figure 2. (a) Original image (b) stego with 1 KB (c) stego with 2 KB (d) stego with 4 KB (e) stego with 6 KB (f) stego with 8 KB (g) stego with 10 KB**

Clearly, according to figure (2), there are no noticeable dissimilarities between the stego images (b) to (f) with the original image. Hence, this is robust evidence that the proposed method does not highly affected when the secret file size increased.

Moreover, the results of the PSNR for different file sizes were demonstrated in table (6). With regard to these results, the values of the PSNR are very sophisticated. In the language of numbers, 10 KB of text with nearly 44 (dB) PSNR is really excellent.

**Table 6. PSNR (dB) for test images**

| Text file size | Lena | Saif | Peppers |
|---|---|---|---|
| 1 KB | 44.6920 | 44.8646 | 44.3272 |
| 2 KB | 44.6086 | 44.7809 | 44.2457 |
| 4 KB | 44.4492 | 44.6199 | 44.0901 |
| 6 KB | 44.3091 | 44.4826 | 43.9453 |
| 8 KB | 44.1522 | 44.3217 | 43.7805 |
| 10 KB | 44.0073 | 44.1803 | 43.6396 |

## 6. COCLUSIONS

In this paper, a novel method for steganography based on the FMM method has been proposed. Many researchers have been reported different techniques but all the methods suffer with image quality problem. So, in order to achieve good quality, the implementation of the FMM into steganography produces better results that do not have a noticeable distortion on it by the human eye. The stego images were also tested using PSNR value. According to the PSNR value, the stego images have high PSNR. Hence, ST-FMM novel steganography algorithm is very efficient to hide the text inside the image. ST-FMM is not an absolute steganographic algorithm and has some limitations. There are issues that need to be resolved. One of these issues is that the 5×5 window size is large to accommodate one secret letter. According to eq.(1), when using all of the 256 ASCII characters an 8×8 window size will be used.